\documentclass[showpacs,prl,superscriptaddress,twocolumn]{revtex4}

\usepackage {graphicx}
\usepackage {hyperref}

\begin{document}

\title{Direct Observation of High-Temperature Polaronic Behavior 
In Colossal Magnetoresistive Manganites
}

\begin{abstract}
The temperature dependence of the electronic and atomic structure of the colossal magnetoresistive oxides 
$\mathrm{La_{1-x}Sr_{x}MnO_{3} (x = 0.3, 0.4) }$
has been studied using core and valence level photoemission, x-ray absorption and emission, 
and extended x-ray absorption fine structure spectroscopy. A dramatic and reversible change of the electronic 
structure is observed on crossing the Curie temperature, including charge localization and spin moment increase of Mn, 
together with Jahn-Teller distortions, both signatures of polaron formation. 
Our data are also consistent with a phase-separation scenario.
\end{abstract}

\author{N. Mannella} 
\email[Correspondence and requests for materials should be addressed to N. Mannella: ]{  norman@electron.lbl.gov}
\affiliation{Department of Physics, UC Davis, Davis, CA, USA}
\affiliation{Materials Sciences Division, Lawrence Berkeley National Laboratory, Berkeley, CA, USA}

\author{A. Rosenhahn}
\affiliation{Materials Sciences Division, Lawrence Berkeley National Laboratory, Berkeley, CA, USA}

\author{C. H. Booth}
\affiliation{Chemical Sciences Division, Lawrence Berkeley National Laboratory, Berkeley, CA, USA}

\author{S. Marchesini}
\affiliation{Materials Sciences Division, Lawrence Berkeley National Laboratory, Berkeley, CA, USA}
\author{B. S. Mun}
\affiliation{Department of Physics, UC Davis, Davis, CA, USA}
\affiliation{Materials Sciences Division, Lawrence Berkeley National Laboratory, Berkeley, CA, USA}
\author{S.-H. Yang} 
\affiliation{Materials Sciences Division, Lawrence Berkeley National Laboratory, Berkeley, CA, USA}
\author{K. Ibrahim}
\affiliation{Materials Sciences Division, Lawrence Berkeley National Laboratory, Berkeley, CA, USA}
\affiliation{Beijing Synchrotron Radiation Laboratory, Beijing, China}

\author{Y. Tomioka} 
\affiliation{Correlated Electron Research Center and Joint Research Center for Atom Technology, Tsukuba, Japan}

\author{C. S. Fadley}
\affiliation{Department of Physics, UC Davis, Davis, CA, USA}
\affiliation{Materials Sciences Division, Lawrence Berkeley National Laboratory, Berkeley, CA, USA}

\pacs{71.38.-k  71.30.+h  75.47.Gk  79.60.-i}

\date{\today}
\maketitle

The colossal magnetoresistive (CMR) manganites are among the most studied 
materials in condensed matter physics \cite{*To:2000}, with considerable 
potential for technological applications. And yet agreement on their correct 
theoretical description is still lacking. Although the long-standing 
double-exchange (DE) model \cite{*To:2000,Zener:2001} 
provides a qualitatively correct description of these materials, more 
complex, but not necessarily mutually exclusive, mechanisms such as charge 
localization via Jahn-Teller distortion (JTD) with polaron formation 
\cite{Millis:1996,Teresa:1997,Booth:1998,Louca:1999} 
and phase separation \cite{Moreo:2003} have also been proposed.

In particular, manganites with composition La$_{1 - x}$Sr$_{x}$MnO$_{3}$ 
(LSMO) are still a matter of controversy: they have commonly been referred 
to as ``canonical" (i.e. capable of description via the DE model alone) by 
some authors \cite{*To:2000,Zener:2001}, but have been 
suggested by others to exhibit more complex behavior due to JTDs and polaron 
formation \cite{Millis:1996,Louca:1999}.

We have thus studied the ferromagnetic-to-paramagnetic transition in the CMR 
compounds La$_{1 - x}$Sr$_{x}$MnO$_{3}$ (LSMO, x = 0.3, 0.4) by means of 
temperature-dependent core and valence level photoemission spectroscopy 
(PS), x-ray absorption and x-ray emission spectroscopy (XAS and XES, 
respectively) and extended x-ray absorption fine structure (EXAFS). Our 
measurements reveal significant charge localization onto the Mn atom at high 
temperature, coupled with local JTD, thus providing a direct observation of 
lattice polaron formation. These results thus suggest that the presence of 
polarons above the Curie temperature (T$_{C})$ is a general defining 
characteristic of all the CMR materials.

The LSMO compounds studied are metallic and in a rhombohedral crystal 
structure for the full temperature range accessed by our measurements ($110  
\le  T  \le  600$ K) \cite{Urushibara:1997}. A magnetic phase 
transition from ferromagnetic to paramagnetic metal occurs at T$_{C} \quad  
\approx  370$ K. The PS, XAS and XES spectra have been measured on undulator 
beamline 4.0.2 at the Berkeley Advanced Light Source (ALS) on high-quality 
single crystals which have been fractured \textit{in situ} at room temperature in ultrahigh 
vacuum (ca. 1-2 x $10^{ - 10}$ torr). A quantitative PS analysis of 
core-level photoemission intensities confirmed the expected stoichiometries 
to within experimental accuracy and further showed minimal degrees of 
surface stoichiometry alteration (within $\pm $5{\%} for all species) or 
surface contamination (e.g. less than 0.07 monolayers of C contaminant) 
during the entire length of a given set of experiments (ca. 24 hours). 
Bulk-sensitive EXAFS measurements were carried out on a polished 
single-crystal surface at the Stanford Synchrotron Radiation Laboratory 
(SSRL) on BL 4-1.

\begin{figure}[htbp]
\centerline{
\includegraphics[width=3.05in]{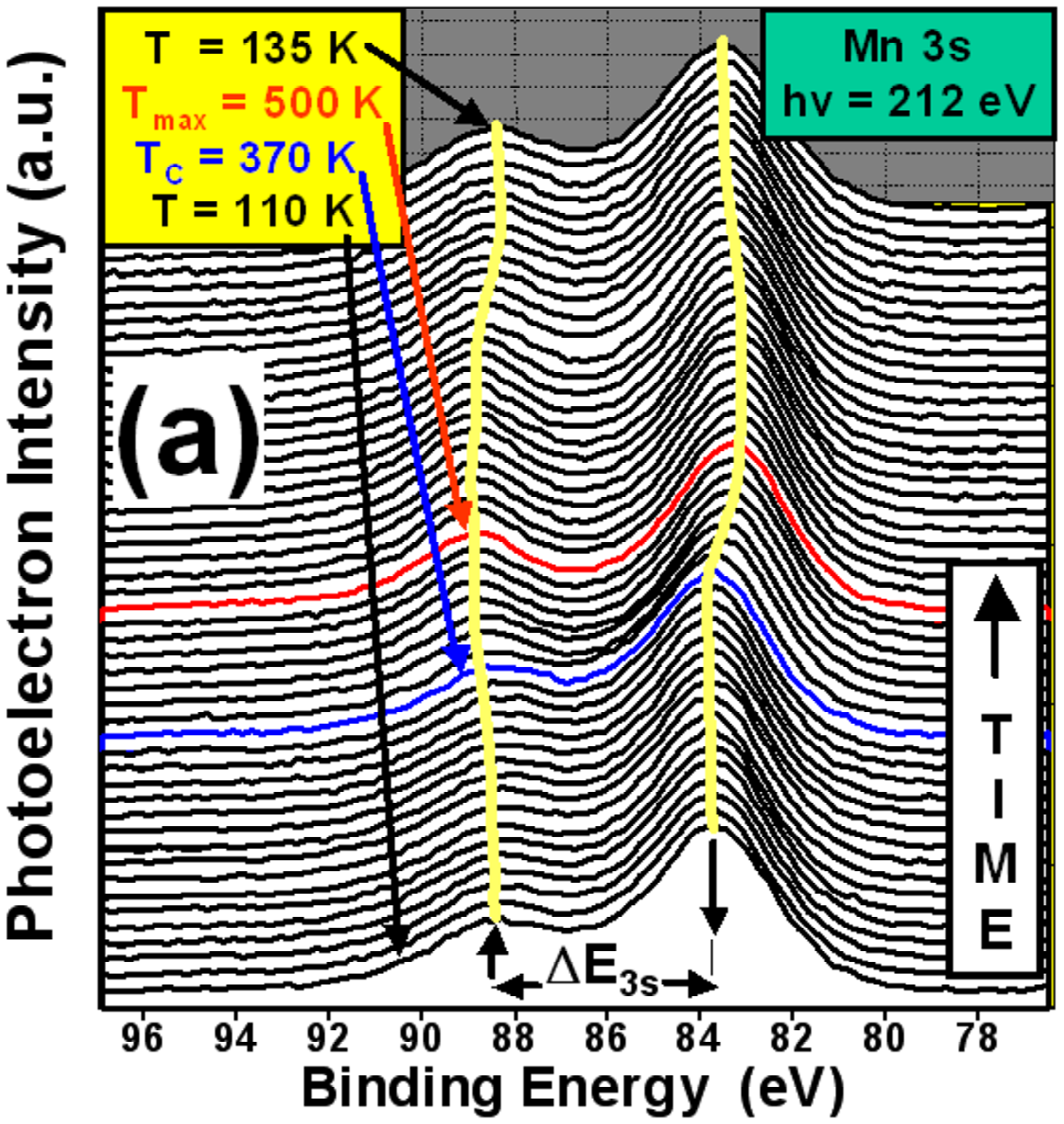}}
\centerline{
\includegraphics[width=3in]{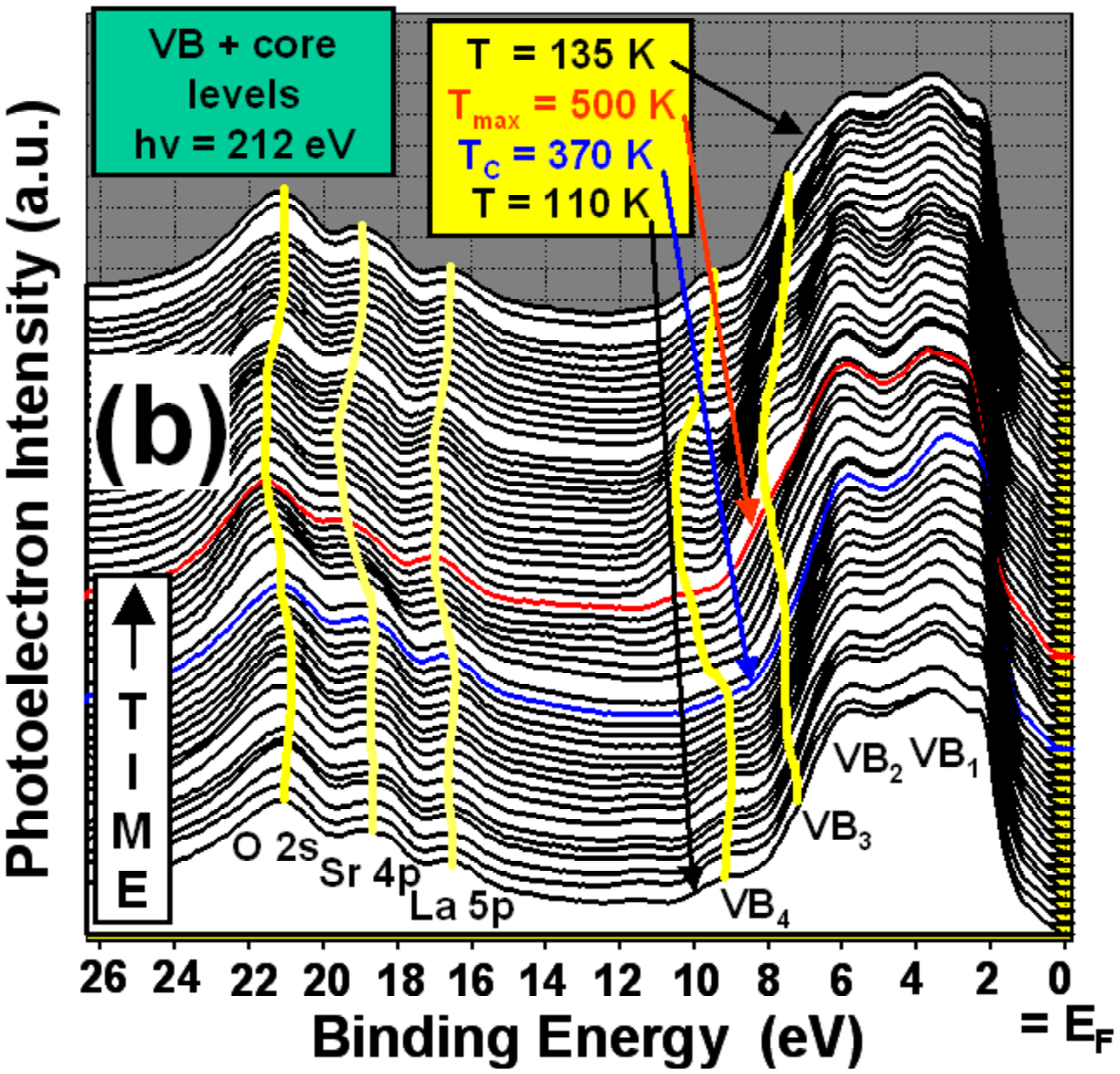}
}
\caption{
Temperature dependent core and valence photoemission spectra from a 
La$_{0.7}$Sr$_{0.3}$MnO$_{3}$ single-crystal. The photon energy was set to 
212.5 eV for the Mn3s spectra (1a) and the high-lying core and valence band 
spectra (1b), and 652 eV for O1s spectra (results summarized in Fig. 
\ref{fig2}b), so 
that the O 1s photoelectrons have very nearly the same kinetic energy (and 
thus escape depth, $\Lambda _{e } \approx $ 5 {\AA}) as the Mn 3s 
electrons. The temperature range was 110 K - 500 K, with a step of 
approximately 20 K. (a) Mn 3s core-level spectra (b) High-lying core-level 
(O 2s, Sr 4p, La 5p) and valence level spectra. Note core-level shifts at 
high T, reversible changes in valence features VB$_{1}$-VB$_{4}$, and lack 
of any Fermi level shift.
}
\label{fig1}
\end{figure}

\begin{figure}[htbp]
\centerline{\includegraphics[width=3.5in]{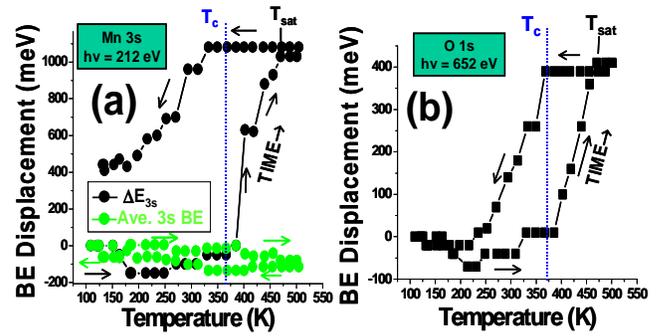}}
\caption{
Reversibility and hysteresis in the temperature dependence of (a) Mn 
3s and (b) O 1s core level binding energies of 
La$_{0.7}$Sr$_{0.3}$MnO$_{3}$, including the Mn 3s multiplet splitting. 
}
\label{fig2}
\end{figure}

We have first detected temperature-dependent electronic structure changes by 
means of core and valence PS (Figs. \ref{fig1} and \ref{fig2}), with the core spectra 
providing an element-specific look at electronic and magnetic states. In 
Fig. \ref{fig1}a, we show a set of Mn 3s photoelectron spectra obtained in a single 
temperature scan on a sample of composition La$_{0.7}$Sr$_{0.3}$MnO$_{3}$. 
These spectra exhibit a doublet due to multiplet splitting of the binding 
energy (BE), a well-known effect in transition metals which provides a 
unique probe of the local spin moment of magnetic atoms 
\cite{Fadley:1969,Gweon:1995}, and which has recently been 
analyzed for several CMR materials \cite{Galakhov:2002}. The multiplet 
energy separation $\Delta $E$_{3s}$ depends on the net spin S$_{v}$ of the 
emitter (Mn in this case) via $\Delta $E$_{3s}$ = (2S$_{v}$ + 
1)J$^{\text{eff}}_{3s - 3d}$, where J$^{\text{eff}}_{3s - 3d}$ denotes the effective 
exchange integral between the 3s and the 3d shells after allowing for 
final-state intra-shell correlation effects 
\cite{Fadley:1969,Gweon:1995,This:1}. The Mn3s 
splitting changes markedly from 4.50 to 5.55 eV as the temperature is raised 
from T$_{C}$ to a higher ``saturation" temperature T$_{\text{sat}} \quad  \approx $ 470 
K beyond which no change is observed. The splitting also tends to return to 
its original value upon cooling the sample (Figs. \ref{fig1}a and \ref{fig2}a), but with 
hysteresis. These data thus indicate a significant increase in the Mn spin 
moment S$_{v}$ at high temperature. Prior results for a range of inorganic 
Mn compounds yield J$^{\text{eff}}_{3s,3d} \quad  \approx $ 1.1 eV 
\cite{Fadley:1969,Gweon:1995,Galakhov:2002,Hermsmeier:1993,Since:1}. 
Using this result, the change in the Mn3s splitting yields an increase in 
its average spin moment from $ \approx $ 3 to $ \approx $ 4 $\mu _{B}$, 
corresponding to about 1 electron transferred to the Mn atom. Note also that 
the average position of the Mn 3s doublet in Fig. \ref{fig1}a changes very little 
with temperature (Fig. \ref{fig2}a).

Simultaneous with measuring the Mn 3s spectra, other core-level (O 1s, O 2s, 
La 4d, La 5p, Sr 4p) and valence-band (VB) spectra were also recorded, with 
all but O 1s and La 4d being shown in Fig. \ref{fig1}b. The BEs of all of the O, La, 
and Sr core peaks are observed to increase by between 0.4 eV and 0.6 eV as 
the temperature changes from T$_{C}$ to T$_{\text{sat}}$, concomitant with the 
change of the Mn 3s splitting. The high-T increases of the core BEs for the 
O, La, and Sr core levels thus suggest charge transfer from the O, La and Sr 
atoms to the Mn atom, fully consistent with the increase of the Mn 3s 
splitting. If charge from the O, La and Sr atoms is transferred to and 
localized on the Mn atom, La, Sr, and O core electrons will experience a 
more positive environment and therefore be detected at higher BE. The 
absence of any shift in the Fermi level E$_{F}$ (to within approximately 50 
meV) also indicates that the observed increase of the core levels BEs is not 
an artifact due to sample charging.

On cooling the sample in the same stepwise fashion, both the Mn 3s splitting 
and the core BEs return to their original values, but with $ \approx $ 
200-K-wide hysteresis loops centered around T$_{C}$, and with a time 
constant of several hours (Fig. \ref{fig2}). The BE loops for O, Sr, and La close on 
the time scale of our measurement, whereas the Mn splitting was found to 
close on a slightly longer time scale. We attribute this difference to 
additional collective magnetic character in the Mn 3s spectrum, perhaps due 
to ferromagnetic cluster formation, resulting in a slower time scale.

We have also explored the depth distribution of these effects below the 
surface. We have observed the same temperature-dependent effects when the 
core spectra were excited with higher photon energies so as to change the 
photoelectron escape depth $\Lambda _{e}$ from $ \approx $ 5 {\AA} (as in 
the data of Figs. \ref{fig1} and \ref{fig2}) to $ \approx $ 15 {\AA} (corresponding to an 
average emission depth of roughly 3 unit cells), indicating that the 
observed electronic structure changes affect more than the outermost surface 
layers. Finally, although not shown here, more bulk sensitive XAS spectra 
measured over the O K- and Mn L-edges and detected with secondary electrons 
of $ \approx $ 100 eV kinetic energy as well as photons in the 
fluorescence-yield mode show remarkable changes when the temperature is 
varied through T$_{C}$ and up to T$_{\text{sat}}$. And the most bulk sensitive XES 
data reveal a strong temperature dependence of the O 2p $ \to $ O 1s and Mn 
3d $ \to $ Mn 2p transitions on crossing T$_{C}$ and approaching T$_{\text{sat}}$. 
Thus, we estimate that these effects take place over at least
the first 30-50 {\AA} inward from the surface, or roughly 6-10 unit cells, 
and are indeed likely to be bulk phenomena. Although all of the results 
presented here are for La$_{0.7}$Sr$_{0.3}$MnO$_{3}$, we have also seen 
identical effects in experiments on La$_{0.6}$Sr$_{0.4}$MnO$_{3}$, with the 
only difference being that T$_{\text{sat}}$ does not occur until about 150 K above 
T$_{C}$.

We now ask whether these dramatic electronic structure changes are 
accompanied by local JTDs of atomic positions in the oxygen octahedra 
surrounding each Mn atom, often considered a structural signature of polaron 
formation. Several prior experimental atomic structure studies have provided 
such evidence for the existence of polarons in ``non-canonical'' manganites, 
suggesting that the existence of polarons is directly related to JTDs 
\cite{Teresa:1997,Booth:1998}, but we now consider the 
presence of JTDs in these ``canonical'' compounds. As a first point, the 
long-time-scale hysteresis of the temperature-driven change of the 
electronic structure (Fig. \ref{fig2}) qualitatively suggests the presence of 
slower-relaxing atomic displacements that are in turn responsible for the 
perturbations of the core and valence levels we observe. However, no obvious 
modification of the long-range crystal structure is shown by x-ray 
crystallography, since our samples are rhombohedral at all temperatures 
studied \cite{Urushibara:1997}. We have thus explored the possibility of 
short-range modifications in structure on passing T$_{C}$ via bulk-sensitive 
Mn $K$-edge EXAFS measurements performed on a La$_{0.7}$Sr$_{0.3}$MnO$_{3}$ 
sample in the fluorescence-yield mode.

\begin{figure}[htbp]
\centerline{\includegraphics[width=3.2in]{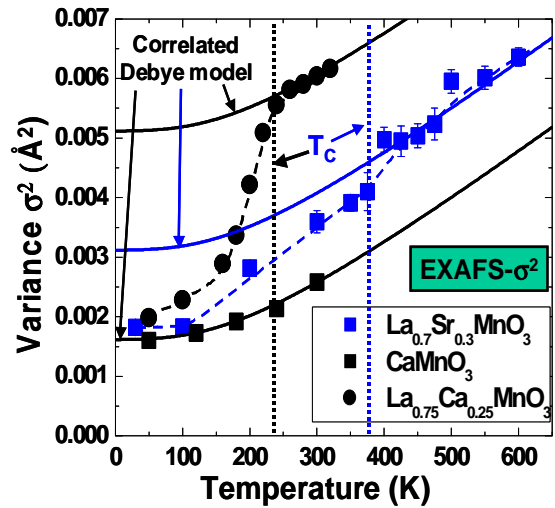}}
\caption{
EXAFS results for the variance of the Mn-O bond length distribution 
$\sigma ^{2}$ with temperature for La$_{0.7}$Sr$_{0.3}$MnO$_{3}$, compared 
with similar data from ref. \cite{Booth:1998} for CaMnO$_{3}$ and 
La$_{0.75}$Ca$_{0.25}$MnO$_{3}$. The solid line through the CaMnO$_{3}$ data 
is a correlated-Debye fit with $\Theta _{cD }$= 860 K. Dashed curves 
represent the experimental data. The same curve is drawn through the 
La$_{0.7}$Sr$_{0.3}$MnO$_{3}$ data except with a 0.0015 {\AA}$^{2}$ offset, 
and again through the La$_{0.75}$Ca$_{0.25}$MnO$_{3}$ data with a 0.0035 
{\AA}$^{2}$ offset. Curie temperatures of La$_{0.75}$Ca$_{0.25}$MnO$_{3}$ 
and La$_{0.7}$Sr$_{0.3}$MnO$_{3}$ are indicated.
}
\label{fig3}
\end{figure}

The EXAFS data permit extracting the temperature dependence of the variance 
of the Mn-O bond length distribution, $\sigma ^{2}$ (Fig. \ref{fig3}). Also shown 
for comparison are some data on the La$_{1 - x}$Ca$_{x}$MnO$_{3}$ series 
from Ref. \cite{Booth:1998}. At low temperatures, the $\sigma ^{2}$ of 
La$_{0.7}$Sr$_{0.3}$MnO$_{3}$ approaches that of the JTD-free CaMnO$_{3}$ 
material. As the temperature is increased toward T$_{C}$, $\sigma ^{2}$ 
increases more rapidly than one expects from a purely vibrational analysis 
via the correlated-Debye model \cite{Crozier:1988}. Above T$_{C}$, 
$\sigma ^{2 }$ increases more gradually, again consistent with a 
Debye-model broadening. These data thus suggest a JTD that develops with 
increasing temperature and saturates once the system becomes paramagnetic, 
as observed in a previous EXAFS study of La$_{1 - x}$Ca$_{x}$MnO$_{3}$ 
\cite{Booth:1998}. The size of the apparent JTD in our Sr-doped 
compound is only about half that in the Ca-doped compound, a difference 
which may be the microscopic cause of the metallic state that survives in 
the paramagnetic state of the Sr-doped (but not Ca-doped) materials.

Our data thus provide direct experimental evidence for the presence of 
polarons in the LSMO compounds, since they show both charge localization 
onto the Mn atom and JTDs. There is theoretical support for this, since 
Millis et al. predicted that, for $x > 0.2$ in LSMO, strong electron-phonon 
coupling should localize the electrons via polarons, with the JT energy 
remaining important even in the metallic state \cite{Millis:1996}. It 
may at first sight seem that polaronic electron localization and metallic 
conduction (our samples are metallic for all temperatures) are mutually 
exclusive, but there is experimental evidence suggesting that in fact they 
are not. E.g., neutron scattering measurements with pair distribution 
function analysis (PDF) have shown that for LSMO up to x = 0.4 local JTDs 
are observed even in the metallic phase \cite{Louca:1999}, with 
somewhat larger magnitudes than reported here. However, our results are in 
quantitative agreement with another recent EXAFS study in which a distortion 
in the metallic phase is observed but to a lesser degree than in the PDF 
analysis \cite{Shibata:2003}. 

It is also possible, and consistent with our data, that the polarons have a 
magnetic character, i.e. that above T$_{C}$ the carriers become localized as 
the lattice is distorted and magnetically polarize the neighboring Mn atoms, 
forming ferromagnetic clusters. Experimental evidence for the existence of 
such ``magnetic polarons" has in fact been reported for a related class of 
compounds \cite{Teresa:1997}. Recent theoretical studies also predict 
the presence of ferromagnetic clusters in the temperature range T$_{C} \quad  
\le $ T $ \le $ T*, with T* adding a new temperature scale, and these 
clusters growing in size when the temperature is reduced from T* to below 
T$_{C}$ \cite{Burgy:2001}. All of our results are consistent with this 
scenario as well, and it is possible that the temperature T* could be 
identified with our temperature T$_{\text{sat}}$.

\begin{figure}[htbp]
\centerline{\includegraphics[width=3in]{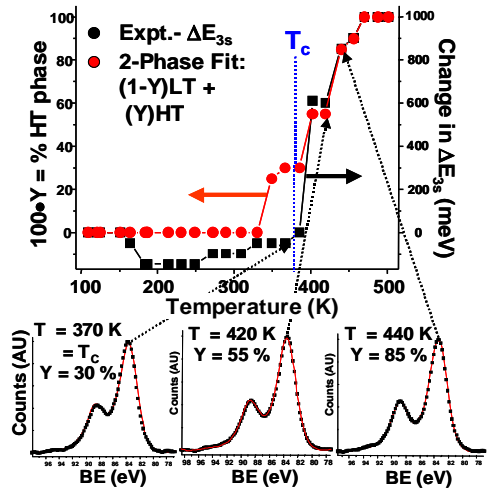}}
\caption{
Two-phase fits of the Mn 3s temperature-dependent core spectra of 
Fig. 1a with a linear superposition of the spectra at low ($T < 200$ K) and 
high temperatures (T $ \ge $ T$_{\text{sat}})$. The fraction Y of the high 
temperature spectrum (red curve in top panel) is the only free parameter. 
The variation of the multiplet splitting $\Delta $E$_{3s}$ (black points and 
curve in top panel) is shown for comparison. Fits to spectra at three 
specific temperatures are shown in the lower panels. 
}
\label{fig4}
\end{figure}

Finally, we comment on the possibility of phase separation in the LSMO 
system, which has been discussed previously from both experimental and 
theoretical viewpoints 
\cite{Louca:1999,Moreo:2003,Shibata:2002}. In order 
to determine whether the change in electronic structure that we observe is 
consistent with such a scenario, we have fitted the temperature-dependent 
core spectra with a linear superposition of the spectra at low temperature 
($< 200$ K) and high temperature ($ \ge $ T$_{\text{sat}})$. The only free parameter 
used is the fraction Y of the high temperature spectrum. Some of the results 
of this type of fit for the Mn 3s spectra are shown in Fig. \ref{fig4}, where there 
is excellent agreement with experiment. Similar agreement was found for fits 
of the O, La, and Sr core levels (not shown). The spectra taken over 
$ \mathrm{T_{C} < T < T_{\text{sat}}}$ can thus be expressed as a linear combination of spectra 
acquired at low and high temperature, further implying the presence of 
unique low- and high- temperature electronic states and spectroscopic 
behavior that is at least consistent with the presence of phase separation 
in these materials. Since our core-level measurements probe only the 
short-range electronic structure, they do not provide any information on the 
sizes of the domains of these two phases.

This study thus provides unexpected results which act to clarify and unify 
our understanding of the CMR materials. In particular, polaron formation, 
which in LSMO is directly observed via both electronic and atomic structure 
changes, is shown to be an important defining characteristic of the 
high-temperature paramagnetic state of the CMR materials, even when the 
electronic phase is metallic. We suggest that future temperature-dependent 
studies of the type carried out here and involving both core and valence 
level spectroscopies could shed important light on other CMR materials, as 
well as the closely related high-temperature superconductors.

We thank M. West, A. Mei, B. Sell, M. Watanabe, A. Nambu, S.B. Ritchey, E. 
Arenholz, and A.T. Young for assistance with the measurements; Z. Hussain 
for instrumentation development; Y. Tokura for assistance with obtaining 
samples, and A. Cavalleri and M.A. Van Hove for helpful discussions. This 
work was supported by the Director, Office of Science, Office of Basic 
Energy Sciences, Materials Science and Engineering Division, U.S. Department 
of Energy under Contract No. DE-AC03-76SF00098.


\begin{thebibliography}{12}
\bibitem{Since:1}  Since the exchange interaction between the Mn 3s and 3d levels is strongly localized on the Mn atom, we can reasonably assume that the exchange integral J$^{\text{eff}}_{3s - 3d}$ is constant with temperature.
\end{thebibliography}
\end{document}